\documentclass[12pt]{article}

\usepackage{amsfonts,amssymb,amsmath,amscd}
\usepackage[ps,dvips,matrix,arrow,frame,import,curve,color]{xy}
\newcommand{\ra}{\rightarrow}
\newcommand{\bul}{\bullet}

\newcommand{\ZZ}{{\mathbb Z}}
\newcommand{\RR}{{\mathbb R}}
\newcommand{\CC}{{\mathbb C}}

\newcommand{\cL}{{\mathcal L}}

\renewcommand{\part}{\partial}

\newcommand{\op}{\oplus}

\newcommand{\bpartial}{{\bar\partial}}

\newcommand{\cI}{{\mathcal I}}

\newcommand{\ot}{\otimes}

\newcommand{\hF}{{\hat F}}
\renewcommand{\Re}{{\rm Re}}
\renewcommand{\Im}{{\rm Im}}
\newcommand{\cA}{{\mathcal A}}
\newcommand{\cB}{{\mathcal B}}
\newcommand{\tf}{{\tilde f}}
\newcommand{\tg}{{\tilde g}}
\newcommand{\ch}{{\rm ch}}
\newcommand{\cX}{{\check X}}
\newcommand{\hA}{{\hat A}}
\newcommand{\cP}{{\mathcal P}}
\newcommand{\hX}{{\hat X}}

\title{A-branes and Noncommutative Geometry}

\author{Anton Kapustin \\{\small \it California Institute of
Technology, Pasadena, CA 91125, U.S.A.}}

\begin{document}

\begin{titlepage}

\maketitle

\begin{abstract}

We argue that for a certain class of symplectic manifolds the category of
A-branes (which includes the Fukaya category as a full subcategory) is equivalent
to a noncommutative deformation of the category of B-branes (which is equivalent
to the derived category of coherent sheaves) on the same manifold. This equivalence
is different from Mirror Symmetry and arises from the Seiberg-Witten transform which relates
gauge theories on commutative and noncommutative spaces. More generally, we argue that
for certain generalized complex manifolds 
the category of generalized complex branes is equivalent
to a noncommutative deformation of the derived category of coherent sheaves on the same 
manifold. We perform a simple test of our proposal in the case when the manifold in question
is a symplectic torus.

\end{abstract}

\vspace{-6.5in}

\parbox{\linewidth}
{\small\hfill \shortstack{CALT-68-2544}}
\vspace{6.5in}

\end{titlepage}

\section{Introduction and Summary}\label{intro}

It is well-known that to any Calabi-Yau manifold $X$ one can attach two categories of topological
D-branes: the category of A-branes and the category of B-branes. The former is an invariant of
the symplectic structure on $X$, while the latter is an invariant of the complex structure on $X$.
The category of B-branes has been argued to be equivalent to the bounded derived category of coherent
sheaves on the complex manifold $X$, while the category of A-branes is closely related to the Fukaya
category of $X$. These categories play an important role in the mathematical interpretation of
Mirror Symmetry proposed by Maxim Kontsevich~\cite{konts}. 

Our understanding of the category of A-branes is much less satisfactory than that of the category of B-branes. This is true even on the level of objects. Originally it was thought that any A-brane is
isomorphic to a ``twisted complex'' of objects of the Fukaya category, each of which is a Lagrangian submanifold carrying a vector bundle with a flat connections. However, it became clear more recently that on some manifolds there exist A-branes of a different kind (coisotropic A-branes)~\cite{KO}. It is not completely clear at present how to enlarge the Fukaya category to include such objects.\footnote{One possible approach proposed by M.~Kontsevich is to consider the Karoubi envelope of the Fukaya category.}
Another problem is that the definition of morphisms in the Fukaya category is very complicated.

From the point of view of Mirror Symmetry, there is an obvious asymmetry between the definitions of
A-branes and B-branes. The derived category of coherent sheaves is defined in an essentially
algebraic way, while the Fukaya category has a more geometric flavor. This asymmetry makes it difficult to 
establish the equivalence of categories predicted by Mirror Symmetry. For example, it is clear
that the Mirror Transform for tori is closely related to the Fourier-Mukai transform for the derived category of an abelian variety, but it is hard to make this analogy precise. It would be very beneficial
to have a more algebraic description of the category of A-branes, say in terms of modules over some 
algebra or a sheaf of algebras. It was conjectured by many people (see e.g. Ref.~\cite{BS})
that this algebra is related
to the deformation quantization of the symplectic manifold in question, but the precise relation remained
elusive. 

Recently, there was an attempt to make complex and symplectic geometry more similar by regarding them
as special cases of a more general geometry, named ``generalized complex geometry''~\cite{Hitchin,Gualt}.
The role of generalized complex (GC) geometry in physics was discussed in 
Refs.~\cite{Kap} -- \cite{KLi2}.
In particular,
it was argued in Refs.~\cite{Kap,KLi1} that to any ``generalized Calabi-Yau'' one can attach two categories
of ``generalized complex branes.'' The geometry of generalized complex branes has been discussed in 
Refs.~\cite{Kap,Zabz}, while morphisms between them have been studied in Ref.~\cite{KLi2}. It was
also argued in Ref.~\cite{Kap} that for a certain class of generalized complex manifolds the category
of generalized complex branes is equivalent to a noncommutative deformation of the derived category 
of a complex manifold.

In this note we make a step toward a more algebraic description of the category of A-branes for
a certain class of {\it holomorphic symplectic} manifolds. For these manifolds we argue that the 
category of A-branes is equivalent to a noncommutative deformation of the category of B-branes on
the same manifold. We also identify a class of generalized complex (GC) manifolds for which a similar
statement holds. Unlike Mirror Symmetry, this new equivalence relates A-branes and B-branes on {\it the same}
manifold, and does not involve T-duality. In both cases, we make essential use of the proposal
of Ref.~\cite{Kap} concerning the relation between a certain class of generalized complex manifolds
and noncommutative complex manifolds.

Let us formulate our claim concerning the category of A-branes more precisely.
Let $X$ be a compact complex manifold with a holomorphic symplectic form $\Omega$. Let $f$ and $\omega$
be the real and imaginary parts of $\Omega$. We will require $f$ to have periods which are
integral multiples of $2\pi$. In this case we will argue that the category of topological D-branes of type A
for the symplectic manifold $(X,\omega)$ is equivalent to the category of B-branes on a noncommutative deformation of the complex manifold $X$ in the direction of the holomorphic bivector $\Omega^{-1}$. 

The simplest example of a holomorphic symplectic manifold satisfying the conditions
of the previous paragraph is the product of a symplectic torus $(X=V/\Gamma,\omega)$ and its dual 
$\hX=V^\vee/2\pi\Gamma^\vee$ with
the symplectic form $-\omega^{-1}$. The natural complex structure in this case is
$$
J=\begin{pmatrix} 0 & -\omega^{-1} \\ \omega & 0\end{pmatrix},
$$
where we used the obvious integral basis.
The corresponding 2-form $f$ is the curvature of the Poincar\`{e} line bundle on $X\times \hX$.

The outline of the paper is as follows. In section~\ref{GCdef} we provide a brief review of GC structures
and associated categories of topological D-branes. In section~\ref{SW} we give an argument relating
the category of A-branes on a holomorphic symplectic manifold of the kind described above and the
category of B-branes on a noncommutative deformation of the same manifold. We also give a similar
argument for a certain class of GC manifolds and the corresponding categories of GC branes.
In section~\ref{tori} we make our proposal more concrete in the case when $X$ is a torus and perform
a simple test. Additional remarks are collected in section~\ref{concl}.

\section{Generalized complex structures and generalized complex branes}\label{GCdef}

A generalized complex (GC) structure on a smooth manifold $X$ is an endomorphism $\cI$ of the vector bundle
$TX\op TX^\vee$ which squares to $-1$, preserves the obvious pseudoeuclidean metric on $TX\op TX^\vee$,
and is integrable, in a certain sense (see Refs.~\cite{Hitchin,Gualt} for details). Any complex
manifold is a GC manifold, because one can set
\begin{equation}\label{compGC}
\cI(I)=\begin{pmatrix} I & 0\\ 0 & -I^\vee\end{pmatrix},
\end{equation}
where $I$ is the complex structure tensor on $X$.
One can easily check that $\cI$ is a GC structure. A symplectic form $\omega$ on $X$ also gives rise
to a GC structure; one lets
$$
\cI(\omega)=\begin{pmatrix} 0 & -\omega^{-1}\\ \omega & 0\end{pmatrix}.
$$

Given a GC structure $\cI$ and a closed 2-form $B$, one can get another GC structure $\cI'$ by letting
$$
\cI'=M_B^{-1} \cI M_B,
$$
where
$$
M_B=\begin{pmatrix} 1 & 0 \\ B & 1\end{pmatrix}.
$$
This is called a B-field transform. For example, the B-field transform of the GC structure Eq.~(\ref{compGC})
associated with a complex structure $I$ is
$$
\cI(I,B)=\begin{pmatrix} I & 0\\ BI+I^\vee B & -I^\vee\end{pmatrix}.
$$
We will call such GC structures block-lower-triangular. Note that the lower-left block is a 2-form
of type $(2,0)+(0,2)$, which is obtained from $B$ by subtracting its $(1,1)$ component.

In this paper an important role will be played by another class of GC structures which will be called
block-upper-triangular:
$$
\cI=\begin{pmatrix} \alpha & \beta \\ 0 & \gamma\end{pmatrix}.
$$
It follows from the definition of the GC structure that $\alpha$ must be a complex structure on $X$,
$\gamma=-\alpha^\vee$, and $\beta$ must be a Poisson bivector of type $(2,0)+(0,2)$ whose
$(2,0)$ part is holomorphic (with respect to the complex structure $\alpha$). That is, any block-upper-triangular GC structure must have the form
$$
\cI(I,P)=\begin{pmatrix} I & 4\Im\ P \\ 0 & -I^\vee\end{pmatrix}
$$
where $I$ is a complex structure and $P$ is a holomorphic Poisson bivector on $X$ (the factor $4$ was introduced for later convenience).

It has been argued in Refs.~\cite{Kap,KLi1} that to any GC manifold\footnote{More precisely, to any GC manifold admitting a generalized K\"ahler structure, see e.g. Ref.~\cite{KLi1}.}
one can associate a category of
topological D-branes. In the case of the GC structure $\cI(I)$, this category is called the category of B-branes; it is believed that it is equivalent to $D^b(X)$. In the case of the GC structure $\cI(\omega)$,
it is called the category of A-branes. In the case of the GC structure $\cI(I,B)$ it is the derived
category of a gerbe over $X$, where the gerbe is determined by the $(0,2)$ part of $B$. These special
cases are familiar and have been extensively discussed in the literature. It was argued in 
Ref.~\cite{Kap} that in the case of the GC structure $\cI(I,P)$ depending on the holomorphic Poisson
bivector $P$ the category of GC branes is equivalent to the derived category of a noncommutative
deformation of the complex manifold $X$. The deformation is parametrized by the bivector $P$.

We will not repeat all the arguments in favor of the latter equivalence here; it will be sufficient for our purpose to discuss the case of D-branes which are line bundles on a torus $X$. 
Let $\nabla$ be a connection on 
the line bundle, and $F=-i\nabla^2$ be its curvature 2-form. The GC condition requires $F$ to satisfy
$$
FI+I^\vee F=F\beta F.
$$
In the case $\beta=0$, this condition simply says that $F$ is of type $(1,1)$. Thus for a GC structure
$\cI(I)$ a line bundle on $X$ can be regarded as a GC brane only if it is holomorphic. For $\beta\neq 0$,
the condition looks unfamiliar. However, as observed in Ref.~\cite{Kap}, one can bring this condition
to a more familiar form by performing the so-called Seiberg-Witten transform~\cite{SW}. 
The Seiberg-Witten transform
is a 1-to-1 correspondence between gauge-equivalence classes of connections on a commutative torus
and a noncommutative torus of the same dimension. The noncommutativity is controlled by an arbitrary constant bivector $\theta$. In our case, if one chooses $\theta$ to be $2\Re\, P$, the condition on the
curvature $\hF$ of the noncommutative connection turns out to be
$$
\hF I+I^\vee \hF=0.
$$
This means that $\hF$ is of type $(1,1)$, i.e. the line bundle on the noncommutative torus is holomorphic. Thus the Seiberg-Witten transform allows to relate GC branes
for a block-upper-triangular GC structure with B-branes on a noncommutative complex manifold.

Note that one can reformulate the relation between $P$ and $\theta$ as follows: $\theta$ is a real
bivector of type $(2,0)+(0,2)$ whose $(2,0)$ part is equal to $P$. Then $\beta$ and $\theta$ are related
by
$$
\beta=-I\theta-\theta I^\vee,\quad \theta=\frac{1}{2}I\beta.
$$

\section{GC branes and B-branes on noncommutative complex manifolds}\label{SW}

In this section we argue that in some cases the category of GC branes is equivalent to a noncommutative
deformation of the category of B-branes on the same manifold.

\subsection{The argument for the category of A-branes}

Let $(X,\omega)$ be a symplectic manifold. Suppose it admits an A-brane which is a line bundle on $X$
with a unitary connection $\nabla$. According to Ref.~\cite{KO}, the curvature 2-form $f=-i\nabla^2$
must satisfy the following condition:
\begin{equation}\label{abranecond}
f\omega^{-1}f=-\omega.
\end{equation}
Equivalently, one can say that $J=\omega^{-1}F$ must be an almost complex structure on $X$.
One can show that $J$ is automatically integrable~\cite{KO}. Also, it is easy to see that 
$\Omega=f+i\omega$ is of type $(2,0)$; it is also closed and nondegenerate. Thus one can characterize
a symplectic manifold $X$ admitting an A-brane as above in the following way: $X$ must admit a complex
structure $J$ and a holomorphic symplectic form $\Omega$, such that $\Im\ \Omega=\omega$, and the 2-form
$f=\Re\ \Omega$ has periods which are integral multiples of $2\pi$. The last condition ensures
that $f$ can be interpreted as the curvature 2-form of some line bundle.

We are now going to argue that the category of A-branes for such a symplectic manifold is equivalent
to the noncomutative deformation of the category of B-branes on the complex manifold $(X,J)$ in the
direction of the holomorphic bivector $\Omega^{-1}$. The first step is to note that a B-field transform
by a closed 2-form whose periods are integral multiples of $2\pi$ induces an equivalence of the corresponding
categories of GC branes. For sigma-models on a worldsheet without boundaries, this is a well-known statement.
The point is that shifting the B-field by a 2-form $B$ multiplies the integrand in the path-integral
by a factor
$$
\exp i\int_\Sigma \phi^* B
$$
where $\phi$ is a map from the worldsheet $\Sigma$ to the target $X$. Clearly, if $\Sigma$ has no boundaries,
and $B$ has periods in $2\pi\ZZ$, this is a trivial operation which leaves the whole theory unchanged.

If $\Sigma$ has boundaries (i.e. if we consider D-branes), then the above factor is nontrivial. However,
it has the same effect as tensoring all D-branes by a line bundle whose curvature is $B$. Indeed,
this operation mutliplies the integrand by
$$
\exp i\int_{\partial\Sigma} \phi^* A=\exp i\int_\Sigma \phi^*(dA),
$$
where $A$ is the connection 1-form whose curvature is $B$. Clearly, tensoring all D-branes by the
same line bundle has no effect on spaces of open strings, and therefore leave the category of
topological D-branes unchanged.

Now let us apply this statement to the case of the GC structure $\cI(\omega)$, where $(X,\omega)$ is
as above. We take $B=-f$. The transformed GC structure is
$$
\cI'=\begin{pmatrix} \omega^{-1}f & -\omega^{-1} \\ \omega+f\omega^{-1} f & -f\omega^{-1}\end{pmatrix}.
$$
Taking into account Eq.~(\ref{abranecond}), we see that this GC structure is block-upper-triangular: 
$$
\cI'=\begin{pmatrix} J & -\omega^{-1} \\ 0 & -J^\vee\end{pmatrix}.
$$
The corresponding bivector $\theta$ is 
$$
\theta=-\frac{1}{2}\omega^{-1} f\omega^{-1}=\frac{1}{2}f^{-1}.
$$
Its $(2,0)$ part, which controls the noncommutativity of holomorphic coordinates, is
$$
P=\frac{1}{4}(f^{-1}-i\omega^{-1}).
$$
In terms of the holomorphic symplectic form $\Omega=f+i\omega$, this means
$$
P=\Omega^{-1}.
$$

Recalling the proposal of Ref.~\cite{Kap} concerning the mathematical meaning of the category of
GC branes with a block-upper-triangular $\cI$, we conclude that the category of A-branes on $(X,\omega)$
is equivalent to the noncommutative deformation of the derived category of the complex manifold $(X,J)$
by the holomorphic bivector $\Omega^{-1}$. We will make this more precise below in the case when $X$
is a torus.

\subsection{The argument for the category of GC branes}

Suppose now that $(X,\cI)$ is a GC manifold which admits a generalized B-brane which is a line bundle
on $X$ with a unitary connection $\nabla$. We can always write $\cI$ in the block form
$$
\cI=\begin{pmatrix} \alpha & \beta \\ \gamma & -\alpha^\vee \end{pmatrix},
$$
where $\alpha$ is a bundle map from $TX$ to itself, $\beta$ is a Poisson bivector on $X$, and $\gamma$ is a 2-form on $X$ (not necessarily closed).

According to Ref.~\cite{Kap}, the curvature 2-form 
$f=-i\nabla^2$ must satisfy the following condition:
$$
f\alpha+\alpha^\vee f=f\beta f-\gamma.
$$
Let us perform a B-field transform of $\cI$ by letting $B=-f$. The transformed GC structure is
$$
\cI'=\begin{pmatrix} \alpha-\beta f & \beta \\ \gamma+f\alpha+\alpha^\vee f-f\beta f & 
f\beta-\alpha^\vee\end{pmatrix}
$$
We see that $\cI'$ is block-upper triangular, therefore the endomorphism
$$
J=\alpha -\beta f
$$
is a complex structure on $X$, while the bivector $\beta$ must be of type $(2,0)+(0,2)$, and
its $(2,0)$ part is a holomorphic Poisson bivector. 

Since the B-field transform induces an equivalence of categories of topological D-branes, we conclude
that the category of GC branes for the GC manifold $(X,\cI)$ is equivalent to the noncommutative
deformation of the category of B-branes on the complex manifold $(X,J)$, where the noncommutativity
is parametrized by $\beta^{(2,0)}$.

\section{Special case: A-branes on a torus}\label{tori}

\subsection{Noncommutative complex tori}

We argued above that for a certain class of holomorphic symplectic manifolds the category of A-branes
is equivalent to a noncommutative deformation of the category of B-branes. For this result to
be useful, one needs a workable description of B-branes on a noncommutative complex manifold.
In general, it is far from clear how to do this. It follows from the results of Kontsevich~\cite{kontsdef} that for any holomorphic Poisson
bivector there exists a corresponding formal deformation of the category of B-branes. In the present
case the Poisson structure is the inverse of a holomorphic symplectic structure, so one can describe
this deformation fairly explicitly using the Fedosov approach to deformation quantization~\cite{Fedosov}. Namely, according to Ref.~\cite{NT} given a holomorphic symplectic form on a K\"ahler manifold $X$
there is a canonical
deformation of the sheaf of holomorphic functions on $X$ into a sheaf of noncommutative algebras (over formal
power series in the Planck constant $\hbar$). A suitable category of sheaves of modules over this algebra
is a noncommutative deformation of the category of coherent sheaves on $X$, and one can define the
category of noncommutative B-branes to be its derived category, in the usual way. 

Unfortunately, a formal deformation is not sufficient for our purposes. Indeed, expansion in the powers
of the Planck constant is the same as expansion in the powers of the Poisson bivector $P=\Omega^{-1}$;
thus the contribution of holomorphic instantons to the A-model correlators are of order
$$
\exp(-c/\hbar)
$$
for some constant $c$ and are not captured by the formal deformation of the category of B-branes.
In general, it is unclear for which $X$ an actual deformation (i.e. the one capturing terms exponentially small in the ``Planck constant'') exists.

In the case when $X$ is a torus $T^{2n}$ with a constant symplectic form $\omega$, one can do better. It is well-known that for any constant bivector $\theta$ on $T^{2n}$ there exists a deformation
of the algebra of smooth functions on $T^n$ given by the Wigner-Moyal product:
\begin{equation}\label{wignermoyal}
(f\star g)(x)=\int \frac{d^n p}{(2\pi)^n} \frac{d^n q}{(2\pi)^n} \tf(p) \tg(q) e^{-\frac{i}{2} \theta(p,q)} 
e^{-ix(p+q)},
\end{equation}
where
$$
\tf(p)=\int d^n x\  f(x) e^{ipx},\quad \tg(p)=\int d^n x\  g(x) e^{ipx}.
$$
Unfortunately, this formula is nonlocal and therefore it is not possible to define a sheaf of noncommutative
algebras on $T^{2n}$ which would be an actual (i.e. nonformal) deformation of the sheaf of smooth functions
and would agree with Eq.~(\ref{wignermoyal}) on globally-defined functions.
However, to define the category of B-branes on a noncommutative complex torus one can proceed as follows.
First, we go to the universal cover of $T^{2n}$, which is $\RR^{2n}$. Let $\cA$ be the algebra of $C^\infty$ functions on a noncommutative $\RR^{2n}$. That is, it is the Wigner-Moyal deformation of 
$C^\infty(\RR^{2n})$. A vector bundle on a noncommutative
$\RR^{2n}$ is simply a free module of finite rank over the algebra $\cA$ which is the Wigner-Moyal deformation of $C^\infty(\RR^{2n})$. For example, the bundle of differential forms is the tensor product of 
$\cA$ and the exterior algebra of $\RR^{2n}$. This bundle is graded in an obvious way and carries a
differential of degree $1$ (the usual exterior derivative). This particular bundle is actually a differential
graded algebra, which is a noncommutative deformation of the de Rham DG-algebra of $\RR^{2n}$.

Given a complex structure $I$ on $\RR^{2n}$, we can decompose the bundle of differential forms into
$(p,q)$ components $\Omega^{p,q}$. The de Rham differential decomposes as $d=\partial+\bpartial$, where $\partial$ has bidegree $(1,0)$ and $\bpartial$ has bidegree $(0,1)$. The algebra
$$
\Omega^{0,\bul}=\op_{p=0}^n \Omega^{0,p}
$$
equipped with $\bpartial$ is a noncommutative DG-algebra which can be regarded as a deformation of the
Dolbeault DG-algebra of $\CC^n$. We will denote this DG-algebra $\cB$.

A holomorphic structure on a vector bundle $E$ over the noncommutative $\CC^n$ is the structure of
a $\cB$-module on $E\ot_\cA \Omega^{0,\bul}$. That is, it is a linear map
$$
\bpartial_E: E\ot_\cA \Omega^{0,\bul}\ra E\ot_\cA \Omega^{0,\bul +1}
$$
satisfying
$$
\bpartial_E (\alpha \cdot s)=(\bpartial\alpha )\cdot s+(-1)^{|\alpha|} 
\alpha\cdot\bpartial_E s,\ \forall s\in E,\ \forall \alpha\in \Omega^{0,\bul},
$$
as well as
$$
\bpartial_E^2=0.
$$
Since holomorphic vector bundles on $\CC^n$ are a special kind of DG-modules over $\cB$, they form a DG-category. Then we consider the category of twisted complexes
for this DG-category and the corresponding homotopy category, and declare the latter to be the category
of B-branes on the noncommutative $\CC^n$. 

In order to define the category of B-branes on a noncommutative
complex torus $T^{2n}$, we recall that $T^{2n}$ is obtained as a quotient of $\CC^n$ by $\ZZ^{2n}$
and equivariantize everything in sight with respect to this $\ZZ^{2n}$ action. That is, vector bundles
on $\CC^n$ are now equipped with an action of $\ZZ^{2n}$ compatible with the $\cA$-module structure,
and morphisms are required to commute with this action. 

The reason we think this definition of the category of noncommutative B-branes is the correct one
is the following. In the physical approach to B-branes, coherent sheaves do not appear at all. What
usually appears is the Dolbeault resolution of a holomorphic vector bundle (or a complex of holomorphic
of vector bundles). Therefore it is natural to use Dolbeault resolutions in the noncommutative case as well.
A similar approach to the definition of the derived category of a noncommutative complex torus was
advocated by Jonathan Block~\cite{Block}.

Noncommutative tori with complex structures and holomorphic vector bundles on them have been studied recently
by Polishchuk and Schwarz~\cite{PS}. However, in that paper the Poisson bivector is of type $(1,1)$ rather
than of type $(2,0)$. In other words, the holomorphic coordinates still commute, but the antiholomorphic
ones do not commute with the holomorphic ones. The category of B-branes is unaffected by such noncommutativity and coincides with the derived category of the commutative torus.

\subsection{A simple test of the proposal}

In this subsection we perform a simple consistency check on our proposed description of the category
of A-branes on a holomorphic symplectic torus $(X,\omega)$. Suppose that we are given a pair of line bundles $\cL_1,\cL_2$ with connections on such a torus such that the curvatures of the connections satisfy the condition
$$
F\omega^{-1}F=-\omega.
$$
That is, $\cL_1$ and $\cL_2$ are A-branes. We would like to compute the Euler characteristic of the space of morphisms between these two objects of the A-brane category. One simple way to do this is to apply mirror
symmetry and make use of the Hirzebruch-Riemann-Roch theorem. According to this theorem,
the desired Euler characteristic is given by
$$
\langle \ch(E_1), \ch(E_2)\rangle,
$$
where $E_1$ and $E_2$ are mirrors of our two line bundles, $\ch$ denotes the Chern character,
and the angular brackets denote the Mukai pairing (in the even cohomology of the mirror torus $\cX$):
$$
\langle \eta, \xi\rangle=(-1)^p \int_\cX \eta\cdot \xi,\quad \eta\in {\rm H}^{2p}(\cX),\ 
\xi\in {\rm H}^{2q}(\cX).
$$
We can simplify this further by noting that the Chern character transforms very simply under any T-duality
(including mirror symmetry). Suppose $X=A\times B$, where $A$ and $B$ are tori. T-dualizing $A$ gives
us the torus $\cX=\hA\times B$. On $A\times \hA$ we have the Poincar\`{e} line bundle whose Chern
character we denote $\cP$. The Chern character of a brane transforms under T-duality as follows:
$$
\ch_X\mapsto \int_A \ch_X \cdot \cP.
$$
One can easily check that the Mukai pairing is invariant under
T-duality, up to a sign. Thus, up to a sign, the Euler characteristic is also given by the Mukai pairing of the Chern characters of our two line bundles $\cL_1$ and $\cL_2$ on $X$. 

On the other hand, we can try to compute the space of morphisms and its Euler characteristic
by applying the B-field transform and the Seiberg-Witten transform, as described in the preceding
sections. Then one gets a pair of holomorphic line bundles on a noncommutative torus, and the
space of morphisms between them is, by definition, noncommutative $\bpartial$ cohomology.
Although we do not know of a suitable ``noncommutative Hirzebruch-Riemann-Roch theorem,'' we can use
the following trick to compute the Euler characteristic of the space of morphisms. First, we can
reinterpret the computation as the index computation for a twisted Dirac operator on
the noncommutative torus. This works in exactly the same way as in the commutative case.
Then we recall that our noncommutative line bundles were obtained from certain commutative line bundles
by means of the Seiberg-Witten transform with some particular bivector $\theta$. Let us scale
this bivector to zero. In the limit of vanishing noncommutativity, we go back to our original
commutative line bundles $\cL_1$ and $\cL_2$. The Dirac index in this limit is given by the Mukai pairing of their Chern characters, by the usual Atiyah-Singer theorem. On the other hand, the index is an integer, and cannot change under continuous deformations. Thus we conclude that the Euler characteristic of the space of morphisms between noncommutative holomorphic line bundles is equal to the Mukai pairing of
the Chern characters of the corresponding commutative line bundles $\cL_1$ and $\cL_2$. This agrees with the result obtained by means of mirror symmetry.

\section{Concluding remarks}\label{concl}

It would be interesting to further test the proposed equivalence of the category of A-branes
and the category of noncommutative B-branes by computing spaces of morphisms in some special cases.
It would be especially interesting to do this when at least one of the two A-branes is represented by
a Lagrangian submanifold. It is not clear what the analogue of the Seiberg-Witten transform for
Lagrangian A-branes is, but if our proposal is correct, the result should be a complex of noncommutative
holomorphic vector bundles. 

It is amusing to note that the situation considered in this paper ($\omega$ is an imaginary part
of a holomorphic symplectic form whose real part is the curvature of a line bundle) is reminiscent of
the theory of
geometric quantization. In the latter theory, one starts with a symplectic manifold $X$ such that the
periods of the symplectic form are integral multiples of $2\pi$. In our case, as explained above, the symplectic form controlling the noncommutativity is $2f$, where $f$ is the curvature of a line bundle,
and therefore the periods of the symplectic form are integral multiples of $4\pi$. The next step
in the theory of geometric quantization is to find a complex structure with respect to which
$f$ has type $(1,1)$ and then consider holomorphic sections of the corresponding line bundle. It would
be interesting to understand the significance of these manipulations for the category of A-branes.

Recently it was observed that spaces of morphisms between coisotropic A-branes on symplectic tori
have a natural structure of modules over {\it real} noncommutative tori~\cite{AZ}. It would be interesting to understand the precise relation between this result and the observations of the present paper.

\section*{Acknowledgments}
I would like to thank Dima Orlov, Oren Ben-Bassat, Jonathan Block, Tony
Pantev, and Marco Gualtieri for helpful discussions. I am also
grateful to the organizers of the Workshop on Mirror Symmetry 
at the University of Miami for providing a stimulating atmosphere.
This work was supported in part by the DOE
grant DE-FG03-92-ER40701.

\end{document}